\documentclass[]{aa}
\usepackage{graphics} 
\usepackage{rotating}
\usepackage{natbib}
\bibpunct{(}{)}{;}{a}{}{,} 
\newcommand {\msol}{\mbox{M$_{\odot}$}}

\newcommand{\ratioo} {N({\rm H}_2) / I_{\rm CO}}
\newcommand{\kms}   {{\rm \, km \, s^{-1}}}

\def\ga{\lower.5ex\hbox{$\; \buildrel > \over \sim \;$}}
\def\la{\lower.5ex\hbox{$\; \buildrel < \over \sim \;$}}
\begin{document}
\title{A Molecular Gas Bridge between the Taffy Galaxies}
\author{J. Braine\inst{1}, E. Davoust\inst{2}, M. Zhu\inst{3},
Ute Lisenfeld\inst{4}, Christian Motch\inst{5}, E.R. Seaquist\inst{3}}
\offprints{J.~Braine, braine@obs.u-bordeaux1.fr}
\institute{ Observatoire de Bordeaux, UMR 5804, CNRS/INSU, B.P.
89,  F-33270 Floirac, France
\and UMR 5572, Observatoire Midi-Pyr\'en\'ees,11 Avenue Edouard Belin, 
31400 Toulouse, France
\and Univ. of Toronto, Dept. of Astronomy \& Astrophysics, 
60 Saint George St., Toronto, ON M5S 3H8, Canada
\and Instituto de Astrof\'{i}isica de Andaluc\'{i}a, CSIC, Apdo. Correos 3004
18080 Granada Spain
\and Observatoire de Strasbourg, 11 rue de l'Universit\'e, 67000 Strasbourg}
\date{Received / Accepted}
\authorrunning{Braine et al.}
\titlerunning{}
\abstract{
The Taffy Galaxies system, UGC 12914/5, contains huge amounts 
of molecular gas in the bridge region between the receding spirals 
after a direct collision.
$2 - 9 \times 10^9$M$_\odot$ of molecular gas is 
present {\it between} the galaxies, more than the CO emission from the 
entire Milky Way!  Such dense gas can only be torn off by 
collisions between dense clouds, in this case with relative velocities of about 
800 $\kms$, such that the remnant cloud acquires an intermediate velocity
and is left in the bridge after separation of the colliding galaxies.  
We suggest that after ionization in the collision front, the gas
cooled and recombined very quickly such that the density remained high and the 
gas left the colliding disks in molecular form.
\\
\keywords{Galaxies: spiral -- Galaxies: evolution -- Galaxies: ISM -- 
Galaxies: interaction -- Galaxies: individual UGC 12914 -- 
Galaxies: individual UGC 12915}  }

\maketitle

\section{Introduction}

Galaxy interactions are major drivers of galaxy evolution. Outside
of clusters, most interactions are tidal, in which the relative
velocities are similar to or less than the galactic rotation
velocities, and can result in morphological changes and/or galaxy
merging. The other type of interaction is a direct collision at
speeds well beyond galactic rotation velocities such that 
tidal forces do less damage 
than the collision itself \citep{Struck99}.  These ``direct collisions'' are rarer.
In a spiral-spiral direct hit, the stars essentially pass right 
through the disk of the other galaxy because of the 
low likelihood of stellar collisions.  The pressure exerted by gas clouds 
(column density $\la 10^{23}$protons cm$^{-2}$) has no effect on the stars
(column density $\ga 10^{34}$ protons cm$^{-2}$).  The ``damage'' is
mostly caused by collisions of gas clouds in the spiral disks. The
direct hits involving gas-rich galaxies are really ISM-ISM
collisions, in which gas is dragged from the spiral disks into
the space between the two systems.  However, because the gas
mass is a small fraction of the disk mass in spirals, these
collisions tend not to form mergers as the stars pass right
through. Tidal interactions can eject gas (and stars) as well but
in tidal tails rather than a gaseous bridge.

The interacting pair of galaxies, UGC 12914/5, is one of the best
examples of a strong ISM-ISM collision.  \citet{Condon93}, 
hereafter C93, observed
the system with the VLA in the H{\sc i} line and 6 and 20 cm continuum
emission. They found that the bridge between UGC~12914 and UGC~12915
emits strong synchrotron emission, with the magnetic fields
being stretched as the galaxies separate, hence their naming the
system the Taffy Galaxies.  C93 showed that the
collision was indeed close to head-on, with the nucleus of
UGC~12915 passing through the more massive UGC~12914 slightly
North of its center.  The disks are counter-rotating and the
recession velocities are nearly the same, so that the relative
velocities are essentially transverse. Both disks are 
massive judging from the size of the system and the rotation
velocities (260 and 310 $\kms$ for UGC 12915 and 12914 respectively). 
C93 estimate that the collision occurred about 20 Myr ago with a 
transverse velocity of about 600 $\kms$.  Considering the counter-rotation
and the transverse velocity, the speed at which clouds collide is
some 800 $\kms$.

In this work we present $^{12}$CO and $^{13}$CO spectra and
an optical slit spectrum of the Taffy system.
Recent studies have shown that, as predicted by C93, the synchrotron bridge
contains little dust \citep{Jarrett99,Zink00}, presumably because
the shocks have destroyed most of the grains. The extremely
interesting feature is that {\it huge quantities of molecular gas
are present in the bridge region}.  \citet{Smith01}
presented a CO spectrum showing CO in the bridge region.
For consistency with previous work, we assume a distance to the Taffy
galaxies of 60 Mpc, corresponding to H$_0$ = 75 km s$^{-1}$ Mpc$^{-1}$.  

\begin{figure*}[t]
\begin{center}
\includegraphics[angle=0,width=16cm]{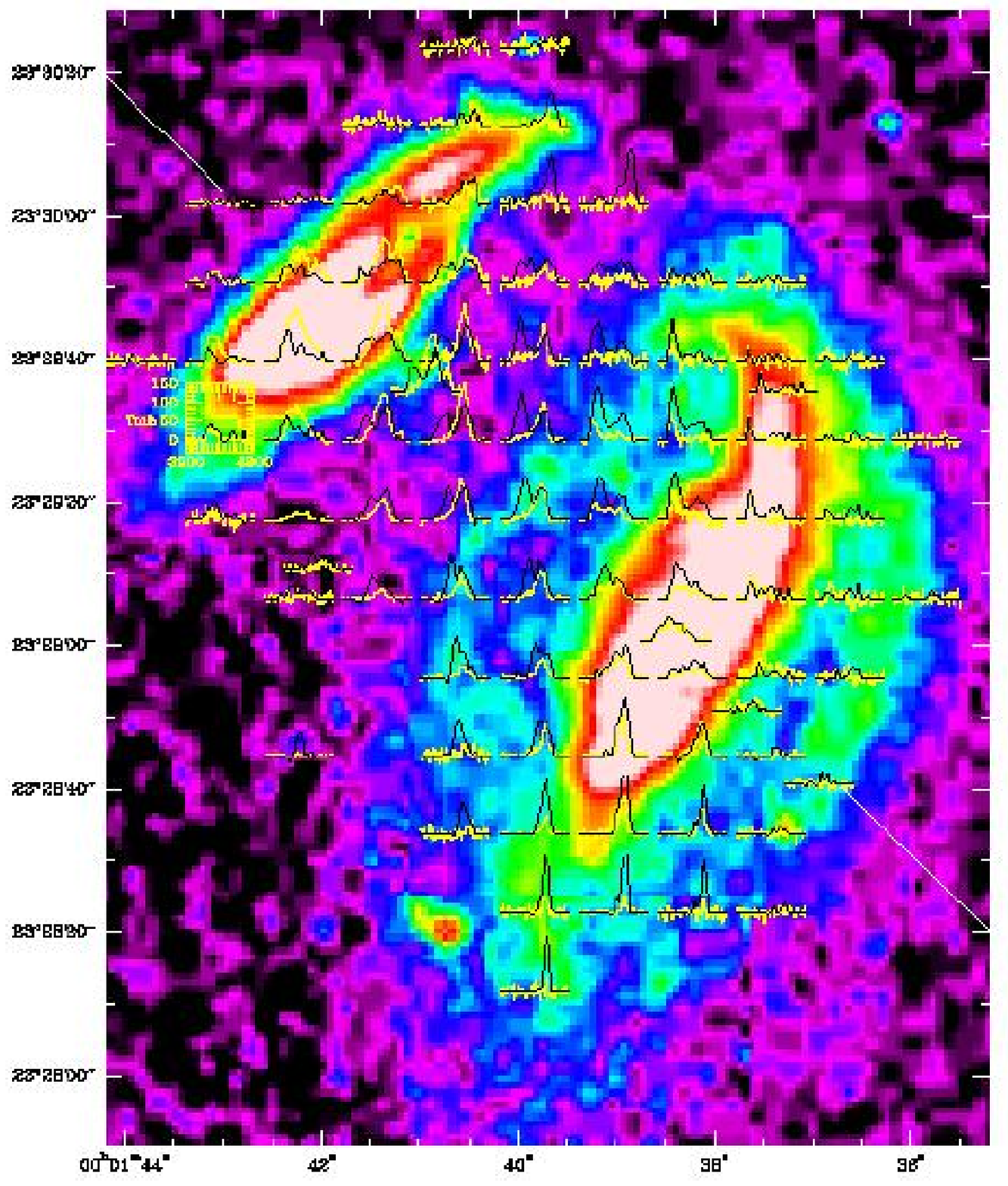}
\caption{CO(1-0) spectra (black line and black scale, intensity
in milliKelvins) overlaid with H{\sc i} spectra (yellow line) on a Digitized Sky
survey image of the UGC 12914/5 system.  The center of UGC 12915 is at
$00^{\rm h}01^{\rm m}41.9^{\rm s}$, $23^\circ29'44.9''$ and the H{\sc ii} region is at
$0^{\rm h}1^{\rm m}40.6^{\rm s}$, $23^\circ 29'33.9''$ and these are the positions
shown in Figure 2 and marked with stars here.  The H{\sc i} emission extends
far south in UGC 12914 (lower right).  Lines mark the ends of the slit 
which roughly connects the nuclei.  The bridge region is delimited by the
white polygon.}
\end{center}
\end{figure*}

\section{Observations}

The millimeter-wave observations were carried out with the 30
meter antenna on Pico Veleta, Spain, operated by the Institut de
RadioAstronomie Millim\'etrique (IRAM).  The data presented here
are from several runs from 1998 to 2001.  All data
are presented using the main beam temperature scale, appropriate
for small sources (see observing details in \citet{Zhu99} and \citet{Braine_tdg2}).
The spatial resolutions (beam size at full width half maximum) are
22$"$ and 11$"$ for respectively the $J = 1 \rightarrow 0$ and $J
= 2 \rightarrow 1$ transitions of $^{12}$CO and $^{13}$CO.

In Figure 1, we present an optical image of the UGC~12914/5 system
with our CO(1--0) spectra (yellow line) and H{\sc i} spectra (black line)
from C93 superposed.

\begin{figure}[t]
\begin{center}
\includegraphics[angle=270,width=8.6cm]{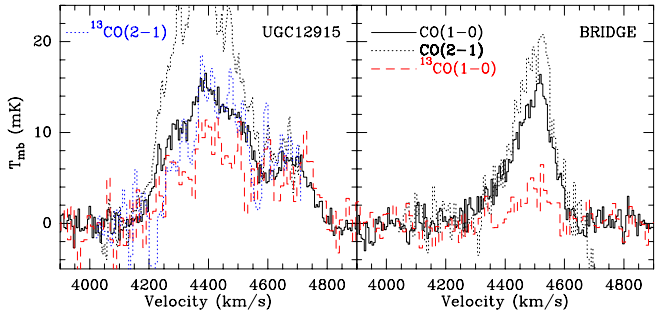}
\caption{$^{12}$CO and $^{13}$CO spectra for UGC~12915 (left) and
the giant H{\sc ii} region (right) in the bridge.  The
$^{12}$CO line intensities are divided by 10.  The angular 
resolutions are respectively 11$''$ and $22''$ for the (2--1) and 
(1--0) transitions. }
\end{center}
\end{figure}

In Figure 2, we present our $^{12}$CO and $^{13}$CO spectra for
the giant H{\sc ii} region in the bridge and for
UGC~12915.  The $^{12/13}$C abundance ratio is expected to be
about 60, so the rarer $^{13}$CO species can be used to estimate
optical depth.  The ratio of the $^{12}$CO to $^{13}$CO line
intensities is very different in the two regions shown.  
In the center of UGC~12915 the
ratio is about 15 (19 in the $J=2 \rightarrow 1$ transition, a
value between the standard 7 -- 10 \citep{Sage91} and the high
values observed  in the very IR-bright galaxies \citep{Casoli92}.
The optical  depth of the CO lines in the bridge is clearly much
lower, as the $^{12}$CO to $^{13}$CO line ratio is about 50 in the
(1-0) transition and $^{13}$CO(2--1) is undetected, such that the
intensity ratio is $\ga 100$. 

The optical long-slit spectra were taken with the Carelec
spectrograph \citep{Lemaitre90} on the 1.93m telescope
of Observatoire de Haute-Provence in November of 2002. The 
CCD was an EEV 2048$\times$1024, with a
pixel size of 13.5$\mu$, which corresponds to
spectral and spatial resolutions of 1.78\AA\ and 0.58" 
respectively. We obtained a 45-minute spectrum of the pair of galaxies
with the slit aligned with the nucleus of UGC 12914 and the giant 
H{\sc ii} region as shown in Fig. 1.
Spectra of standard stars were obtained on a different night.

The main goal of these observations was to determine how the ionized
gas is excited: by young stars from a recent starburst or by
shocks.  The line ratios indicate that the gas in the nucleus of
UGC 12915 and in the giant H{\sc ii} region are photoionized, and that
UGC 12914 is a LINER. The metallicities from O{\sc iii}/H$\beta$
line ratios are subsolar -- $12+$log(O/H)$ = $ 8.5$\pm 0.2$ and 
8.7$\pm 0.3$ for the giant H{\sc ii} region and the nucleus of UGC 
12914 respectively.  Figure 3 shows the optical spectra of
UGC 12914, UGC 12915, and the giant HII region.

\begin{figure}[t]
\begin{center}
\includegraphics[angle=270,width=8.8cm]{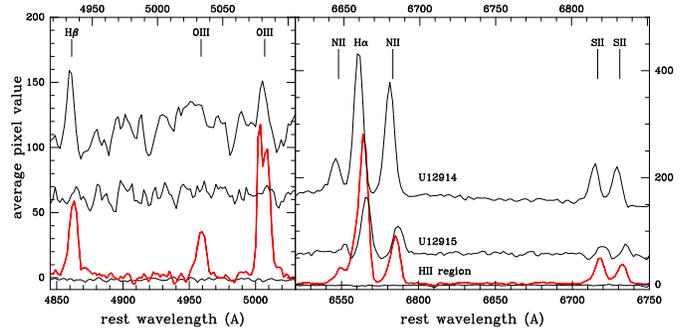}
\caption{Optical spectra of the Taffy system around the H$\beta$ 
and H$\alpha$ lines.  Top line is nucleus of UGC 12914
(intensities divided by 2), followed by UGC 12915 and the giant
H{\sc ii} region (thick red line). No emission is detected (bottom
line) between the H{\sc ii} region and  the outer ring of UGC
12914.  The O{\sc i} 6300\AA\ line is also detected in  UGC 12914
and the H{\sc ii} region and is about 0.3 -- 0.5 as  strong as the
N{\sc ii} 6548\AA\ line.}
\end{center}
\end{figure}



\section{Gas Masses and the bridge region}


Integrating over the entire region observed in CO(1--0) and using
a``standard'' conversion ratio $\ratioo = 2 \times 10^{20}$
cm$^{-2}$ per K km s$^{-1}$ to convert the CO emission into an
H$_2$ column density, our estimate of the total gas mass (H + He)
in the molecular clouds is $3.5 \times 10^{10} \msol$.  25\% of
the CO emission comes from the bridge region, representing more
gas than in the whole of the Milky Way! {\it How can so much
molecular gas be pulled out of a galactic disk and what does this
fact tell us about the ISM?}

\begin{table}
\begin{center}
\begin{tabular}{llll}
 & bridge & U12914 & U12915 \\
\hline
 $\langle$ I$_{\rm CO}$ $\rangle$ & 12.2 & 6.2 & 16.4 \\
 L$_{\rm CO(1\rightarrow 0)}$ &2.0 10$^9$ & 2.7 10$^9$& 3.2 10$^9$ \\
M$_{\rm HI}$ & 2, 5 $\times 10^9$ & 7.5 10$^9$ & 2.7 10$^9$\\
M$_{\rm H_2}$ & 2 -- 9 10$^9$ & 5 -- 10 10$^9$ & 3 -- 6 10$^9$ \\
\end{tabular}
\caption[]{Average CO(1--0) intensities for each region in K
$\kms$, CO(1--0) luminosity in K km s$^{-1}$ pc$^2$, and H{\sc i}
and H$_2$ masses in solar units.  The two values given for the
H{\sc i} mass in the bridge are the H{\sc i} mass in the same
region as the CO (see Sect. 3) and the total bridge H{\sc i} mass.
The H$_2$ mass range takes into account the possible errors in the
$\ratioo$ conversion as discussed in Sect. 3. The Hydrogen masses
include Helium in order to give the total gas mass.}
\end{center}
\end{table}


First, let us address the question of the $\ratioo$ conversion
factor.  The ``standard'' factor used in the calculation above,
$\ratioo = 2 \times 10^{20}$ cm$^{-2}$ per K km s$^{-1}$, is
likely overestimated for galactic nuclei and starburst galaxies.
While neither UGC~12914 nor UGC~12915 are strong starbursts,
UGC~12915 is a strong FIR emitter {\it and} its CO emission is
rather centrally concentrated.  In UGC~12914, the situation is the
reverse -- most of the CO emission comes from the disk, with CO
detected out to distances of 10 -- 12 kpc from the center.  We
tentatively conclude that (a) the true H$_2$ mass of UGC~12915
could be overestimated by up to a factor $\sim 4$ but (b) the
H$_2$ mass of UGC~12914 is unlikely to be grossly overestimated.

{\it How much molecular gas is present in the bridge region?} The 
large amount of molecular gas was not predicted by C93, who expected
that ($a$) the molecular gas was too dense to be pulled out and
($b$) the molecules involved in collisions strong enough to blow a
molecular cloud out of a disk would be shock destroyed. Our
observations show that this is not the case.

As noted in Sect 2, the optical depth of the CO lines is clearly
lower in the bridge because of the high $^{12/13}$CO intensity
ratio (Fig. 2).  As predicted by C93, and attributed to dust
destruction, the FIR emission in the bridge region is weak
\citep{Zink00}. Dust mantles contain a lot of carbon and oxygen
which, when expelled from the grain, can increase the gas-phase CO
abundance. Because the CO emission here is not very optically
thick, increasing the CO abundance can increase the intensity of
the CO emission per H$_2$ mass. The low optical depth and increase
in CO abundance due to grain destruction taken together could
decrease the $\ratioo$ ratio by a factor of a few as compared to
the ``standard'' value.  A more precise estimate of the molecular
gas mass is currently impossible.  Nonetheless, even reducing the
bridge molecular gas by a factor of four means that the bridge
alone contains as much or more H$_2$ as the whole Milky Way!

The geometry of the collision in the Taffy system maximizes the 
cloud collisions and is probably unusually efficient at drawing 
gas out of the disks.  What kind of collision is capable of bringing
some $10^{10}$ M$_\odot$ of gas out of the disks?  Many of the
line profiles (Fig. 1) in the bridge region show double H{\sc i} peaks.
The CO is systematically coincident with the higher velocity HI
peak.
The low-velocity H{\sc i} peak in the bridge with no CO counterpart is
presumably the unperturbed H{\sc i} belonging to the outer northeastern
part of UGC 12914 and located behind the bridge.

One would expect head-on GMC (Giant Molecular Cloud)-GMC 
collisions at $800 \kms$ to result in ionization
because the kinetic energy dissipated is close to 1 keV/proton.
The magnetic field energy is negligible compared to the cloud kinetic 
energy dissipated in collisions so magnetic fields should not be able 
to pull the dense gas out of the disks.

{\it Could much of the molecular gas have recombined in the bridge
region since the collision?}  \citet{Braine_tdg2} estimate the
H$_2$ formation time to be $t_{20\%} \sim 10^7 / n$ years where
$n$ is the atomic hydrogen density in atoms cm$^{-3}$ and
$t_{20\%}$ is the time for 20\% of the H{\sc i} to become H$_2$.  We can
estimate the density by simply taking the hydrogen column density
and dividing by the size of the bridge, likely to be similar in
depth and in width.  Thus, for $N_{\rm H} \approx 10^{22}$
cm$^{-2}$ and $d \approx 10$ kpc, we obtain $n_{\rm H} \approx
0.3$ cm$^{-3}$.  Thus, in the time since collision, we do not
expect more than 20\% 
of the H{\sc i} in the bridge to have recombined
into H$_2$ -- this is a small fraction of the bridge H$_2$ mass.

However, \citet{Harwit87} argue that while GMC-GMC collisions result
in complete ionization, the cooling time for dense (several $10^3$
cm$^{-3}$) gas from $10^6$ to below $10^4$ K is less than 100 years.
In such a short time the cloud volume cannot increase appreciably
so the gas remains dense.  The recombination time scale
is quite short for such dense gas, provided the photons can
escape or be absorbed by dust. If so, then the H$_2$ formation time 
as described above should also be very short due to the high density.
{\it While most H$_2$ was not formed in the bridge, colliding GMCs may 
have been rapidly ionized only to cool and reform H$_2$ while the galactic 
disks are still passing through each other}.  This would result in a H 
recombination line flash for each cloud collision.  Such a mechanism 
could explain the mass of gas spread over the bridge region.
Collisions between diffuse gas clouds will also result in their
ionization but at the lower densities the gas will not have time to
cool and recombine before cloud dissipation, hindering the formation
of the large amounts of H$_2$ observed here.

The other clear case of a recent ISM-ISM collision is the UGC 813/6 
system which also has a synchrotron and H{\sc i} bridge \citep{Condon02}. 
In Stephan's Quintet, in which NGC 7318b is hitting the compact group
at about 800 km s$^{-1}$, \citet{Lisenfeld02} have shown that about 
$3 \times 10^9$M$_\odot$ of H$_2$ is present in the
colliding region SQ A, some 20 kpc from the centers of NGC 7318a and b.
It seems clear that although the velocities should result in
ionization of the neutral gas, ISM-ISM collisions below 1000 km s$^{-1}$
are extremely efficient at bringing gas out of the galaxies, even from
the inner parts.



\begin{acknowledgements} We would like to thank J.J. Condon and
T. Jarrett for the H{\sc i} and radio continuum data.
\end{acknowledgements}

\bibliographystyle{apj}
\bibliography{jb}

\end{document}